\documentstyle[12pt]{article}

\catcode`\@=11
\long\def\@makefntext#1{
\protect\noindent \hbox to 3.2pt {\hskip-.9pt
$^{{\ninerm\@thefnmark}}$\hfil}#1\hfill}        

 \def\@makefnmark{\hbox to 0pt{$^{\@thefnmark}$\hss}}  

\def\ps@myheadings{\let\@mkboth\@gobbletwo
\def\@oddhead{\hbox{}
\rightmark\hfil\ninerm\thepage}
\def\@oddfoot{}\def\@evenhead{\ninerm\thepage\hfil
\leftmark\hbox{}}\def\@evenfoot{}
\def\sectionmark##1{}\def\subsectionmark##1{}}

\newcounter{sectionc}\newcounter{subsectionc}\newcounter{subsubsectionc}
\renewcommand{\section}[1] {\vspace{0.6cm}\addtocounter{sectionc}{1}
\setcounter{subsectionc}{0}\setcounter{subsubsectionc}{0}\noindent
      {\bf\thesectionc. #1}\par\vspace{0.4cm}}
\renewcommand{\subsection}[1] {\vspace{0.6cm}\addtocounter{subsectionc}{1}
      \setcounter{subsubsectionc}{0}\noindent
      {\it\thesectionc.\thesubsectionc. #1}\par\vspace{0.4cm}}
\renewcommand{\subsubsection}[1]
{\vspace{0.6cm}\addtocounter{subsubsectionc}{1} \noindent
{\rm\thesectionc.\thesubsectionc.\thesubsubsectionc.  #1}\par\vspace{0.4cm}}

\newcounter{appendixc}
\newcounter{subappendixc}[appendixc]
\newcounter{subsubappendixc}[subappendixc]

\renewcommand{\appendix}[1] {\vspace{0.6cm}
        \refstepcounter{appendixc}
        \setcounter{figure}{0}
        \setcounter{table}{0}
        \setcounter{equation}{0}
        \renewcommand{\thefigure}{\Alph{appendixc}.\arabic{figure}}
        \renewcommand{\thetable}{\Alph{appendixc}.\arabic{table}}
        \renewcommand{\theappendixc}{\Alph{appendixc}}
        \renewcommand{\theequation}{\Alph{appendixc}.\arabic{equation}}
        \noindent{\bf Appendix \theappendixc #1}\par\vspace{0.4cm}}

\def\abstracts#1{{
      \centering{\begin{minipage}{30pc}\tenrm\baselineskip=12pt\noindent
      \centerline{\tenrm ABSTRACT}\vspace{0.3cm}
      \parindent=0pt #1
      \end{minipage}}\par}}

\renewenvironment{thebibliography}[1]
      {\begin{list}{\arabic{enumi}.}
      {\usecounter{enumi}\setlength{\parsep}{0pt}
\setlength{\leftmargin 1.25cm}{\rightmargin 0pt}
       \setlength{\itemsep}{0pt} \settowidth
      {\labelwidth}{#1.}\sloppy}}{\end{list}}

\newcounter{itemlistc}
\newcounter{romanlistc}
\newcounter{alphlistc}
\newcounter{arabiclistc}

\newcommand{\fcaption}[1]{
        \refstepcounter{figure}
        \setbox\@tempboxa = \hbox{\tenrm Fig.~\thefigure. #1}
        \ifdim \wd\@tempboxa > 6in
           {\begin{center}
        \parbox{6in}{\tenrm\baselineskip=12pt Fig.~\thefigure. #1}
            \end{center}}
        \else
             {\begin{center}
             {\tenrm Fig.~\thefigure. #1}
              \end{center}}
        \fi}

\newcommand{\tcaption}[1]{
        \refstepcounter{table}
        \setbox\@tempboxa = \hbox{\tenrm Table~\thetable. #1}
        \ifdim \wd\@tempboxa > 6in
           {\begin{center}
        \parbox{6in}{\tenrm\baselineskip=12pt Table~\thetable. #1}
            \end{center}}
        \else
             {\begin{center}
             {\tenrm Table~\thetable. #1}
              \end{center}}
        \fi}

\def\@citex[#1]#2{\if@filesw\immediate\write\@auxout
      {\string\citation{#2}}\fi
\def\@citea{}\@cite{\@for\@citeb:=#2\do
      {\@citea\def\@citea{,}\@ifundefined
      {b@\@citeb}{{\bf ?}\@warning
      {Citation `\@citeb' on page \thepage \space undefined}}
      {\csname b@\@citeb\endcsname}}}{#1}}

\newif\if@cghi
\def\cite{\@cghitrue\@ifnextchar [{\@tempswatrue
      \@citex}{\@tempswafalse\@citex[]}}
\def\citelow{\@cghifalse\@ifnextchar [{\@tempswatrue
      \@citex}{\@tempswafalse\@citex[]}}
\def\@cite#1#2{{$\null^{#1}$\if@tempswa\typeout
      {IJCGA warning: optional citation argument
      ignored: `#2'} \fi}}

\def\fnt#1#2{\footnotetext{\kern-.3em
      {$^{\mbox{\sevenrm #1}}$}{#2}}}

 1
 1
 1

\font\tenbf=cmbx10
\font\tenrm=cmr10
\font\tenit=cmti10

\font\ninerm=cmr9

\renewcommand{\Im}{\mbox{Im}}

\let\a=\alpha \def\b{\bar{\a}}
\let\g=\gamma \let\d=\delta

\def\k{\vec{k}} 
 
\let\x=\xi  \def\q{\vec{q}}
\let\r=\rho \let\s=\sigma 
\let\o=\omega  
  
 \let\S=\Sigma 
  \let\G=\Gamma

\def\2{{1\over2}} \def\4{{1\over4}} \def\52{{5\over2}}
\def\6{\partial }

\def\({\left(} \def\){\right)} \def\<{\langle } \def\>{\rangle }

\def\beg{\begin{equation}}
\def\begar{\begin{eqnarray}}
\def\ee{\end{equation}}
\def\ea{\end{eqnarray}}

\newcommand{\pref}[1]{(\ref{#1})}                
\newcommand{\plabel}[1]{\label{#1}}              
\newcommand{\pcite}[1]{\cite{#1}}                
\newcommand{\pbib}[1]{\bibitem{#1}}              

\input epsf

\newcommand{\ps}{p\!\!\!/}
\newcommand{\pss}{p\!\!\!/\,'}
\begin{document}

\hfill TUW-94-16

\vspace{1cm}

\centerline{\tenbf GAUGE-INDEPENDENT BOUND-STATE CORRECTIONS}
\baselineskip=22pt
\centerline{\tenbf TO THE TOPONIUM DECAY WIDTH\footnote{Presented at the Int.\
Conference \lq\lq Quark Confinement and the Hadron Spectrum", Villa Olmo-Como,
Italy, June 20-24, 1994}}

\vspace{0.8cm}
\centerline{\tenrm WOLFGANG M\"ODRITSCH}
\centerline{\tenrm and}
\centerline{\tenrm WOLFGANG KUMMER}
\baselineskip=13pt
\vspace{0.3cm}
\centerline{\tenit Institut f\"ur Theoretische Physik, Technische
Universit\"at Wien}
\baselineskip=12pt
\centerline{\tenit Wiedner Hauptstra\ss e 8-10, A-1040 Wien,
Austria }
\vspace{0.9cm}
\abstracts{Off-shell and relativistic bound-state corrections for the
decay $t \to b+W$ are calculated to $O(\a_s^2)$ making full use
of the Bethe-Salpeter formalism for weakly bound systems.
Thus we are able to take into account all terms to that order in
a systematic and straightforward manner. One of the previously not
considered contributions cancels precisely gauge dependent terms
which appeared in an earlier off-shell calculation. Important cancellations
also determine the gauge-independent part.}

\vspace{0.8cm}
\rm\baselineskip=14pt

In calculations e.g. of the cross section $\s_{e^+ e^- \to anything}$ near the
$t \bar{t}$ threshold \pcite{Fadin}, it has been realized that
a momentum dependent width (e.g. from phase space reduction) could
lead to drastic changes \pcite{Sumino}.

{\it Qualitatively} the main physical effects from $t$ decaying inside
toponium seem to be quite well understood by now \pcite{kuehn}:
On the one hand, $\G$ decreases by time dilatation and by
reduction of phase space, caused by the decay below the mass shell.
On the other hand, the Coulomb-enhancement induced by the Coulomb interaction
of the relativistic $b$-quark should increase $\G$ as in the muonium system
\pcite{Altar}.

The purpose of our present work is to show that existing quantum
field theoretical technology, well tested e.g. in the (abelian) positronium
case, may be used also here to solve the problem of a gauge-independent
"running width", including the qualitative effects listed above.
Even the Coulomb enhancement can be taken into account properly in this
context, which to the best of our knowledge has not been possible before.
We will use the Bethe-Salpeter (BS) approach as described in \pcite{Kumm1}.

\vspace{.5cm}
The leading imaginary correction to the toponium energy levels
is indicated in figure 1 and gives rise to the BS-perturbation kernel
\beg \plabel{h1}
         H^{\S} = -i [ \Sigma (p) \otimes (\pss-m) + (\ps-m) \otimes  \Sigma
        (p')] (2\pi)^4 \d(k-k'),
\ee
\vspace{.3cm}
where the second term originates from the same contribution of the
$\bar{t}$ quark.

\hspace*{-.8cm}\begin{minipage}[b]{15cm}
   \epsfbox{c: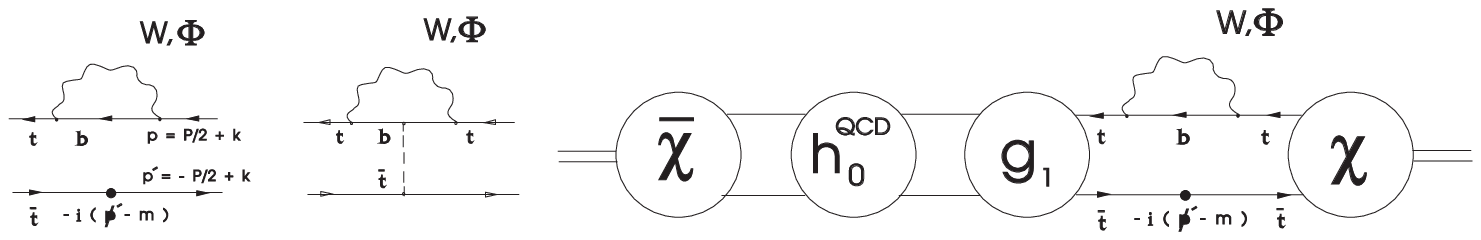}
    \footnotesize
    \hspace*{0.9cm} Fig. 1 \hspace{1.9cm} Fig. 2 \hspace{5.1cm} Fig. 3
    \normalsize
\end{minipage}

\vspace{.3cm}

\noindent
Since we are calculating higher order effects we neglect the mass of
the bottom quark although it may be included in principle.
For the electroweak theory we use the $R_{\xi}$-gauge.
The gauge parameter $\xi$ will not be fixed in the following.

The total contribution of fig.1  to the decay width between Barbieri-Remiddi
(BR) wave functions ( which are our zero order solutions of the BS-equation)
becomes:
\begar \plabel{Gam1}
 \G_1 &=& \G_0 - 2 \G_0 \s_n^2 + \frac{e^2m}{16 \pi s^2} [( 1 +2
\frac{M^2}{m^2} )(1- \frac{M^2}{m^2}) + 2 \r(m^2,\xi)] \< \frac{p^2 -m^2}{m^2}
\>, \ea \begar \r(p^2,\xi)&=& \frac{1-\x}{2\x} ( 1-
\frac{M^2}{2p^2}\frac{1+\x}{\x}) \quad , \qquad \s_n = \frac{2 \a_s}{3 n} \ea
where $\G_0$ is the well known zero order result which is twice the decay
width of a free top quark.  It proves sufficient to evaluate the graphs shown
in fig. 2 to leading order.  We obtain ($ \xi > M^2/m^2$ and $m^2 > M^2$):
\begar
  \G_2 = \frac{e^2}{16 \pi s^2} \< \frac{4\pi\b}{\q^2} \> \left[
(1+\frac{m^2}{2M^2}) (1- \frac{M^2}{m^2})(1+ 3 \frac{M^2}{m^2}) + 4
\r(m^2,\xi) \right] \plabel{Gam2}
\ea
We note already here that with the help of the Schr\"odinger equation the
gauge dependent terms proportional $\r(m^2,\xi)$ exactly  cancel in
\pref{Gam1} plus \pref{Gam2}!

\vspace{.5cm}
also the terms of second order BS-perturbation theory contribute ($h_0^{\S} =
H^{\S} |_{P_0=M_{n,0}}, h_1 =  \6 H/ \6 P_0 |_{P_0=M_{n,0}} $)
\begar
 \G_3 &=& -2 \Im [ \< \< h_0^{QCD}  g_1 h_0^{\S}\> \>+ \< \< h_0^{\S} g_1
h_0^{QCD} \> \>]  \plabel{Gam3}  \\ \G_4 &=& -2 \Im [ \< \< h_0^{QCD}\> \> \<
\< h_1^{\S}\> \> + \< \< h_0^{\S}\> \> \< \< h_1^{QCD}\> \> ] \plabel{Gam4}
\ea
where $h_0^{QCD}$ denotes some QCD perturbation and $\<\<..\>\>$ means between
BR wave functions. $g_1$ denotes the reduced Green function for the state
under consideration.  The first contribution to $\G_3$ is shown in fig. 3.
It is sufficient to evaluate \pref{Gam3} and \pref{Gam4} to
leading nonrelativistic order \pcite{Moed}:
\begar  \plabel{Gam3e}
  \G_{3} &=& \frac{\G_0}{2} \< h_0^{QCD} \> \< \frac{1}{\o} \> \\ \G_{4} &=& -
\frac{\G_0}{2}  \< h_0^{QCD} \>   \< \frac{1}{\o} \> -2 [\Im \< \< h_0^{\S} \>
\>] \< \< \frac{\6}{\6 P_0} K_{BR} \big|_{P_0 = M_{n,0}} \> \>.
\ea

$\G_2$ \pref{Gam2} {\it all} gauge dependent terms are found to cancel.
Moreover a striking feature of the sum $\G_1 + \G_2$ is that even the {\it
gauge-independent } terms precisely sum to zero as well, which is by no means
"natural" in view of the explicit expressions.  In a similar manner also in
the sum $\G_3+\G_4$ large cancellations take place so that finally
\beg  \plabel{erg}
  \G_{bound state} = \G_{0}(1-  \2  \s_n^2 + O(\a_s^3)).
\ee
Recalling that $1-  \s_n^2/2 \approx (1 - \< \k^2/m^2 \>)^{1/2} $, the
residual effect may be interpreted as an approximation of the $\g$-factor for
the time dilatation for the weakly bound top.  The origin of that term, the
relativistically generalized Coulomb-kernel $K_{BR}$ gives some credit to this
interpretation.  For the gauge dependent terms in the sum $\G_1+\G_2$ one can
show that the usual QED-like Ward identity, which is valid here within the
expectation value, is responsible for their cancellation.

A reader familiar with phenomenological calculations may wonder why no running
coupling in $\a_s$ or even a more "realistic" potential involving a confining
piece has been used. The simple answer is that any generalization of this type
would completely ruin the strictly perturbative approach and, as a
consequence, also the built-in gauge-independence advocated here.  Any further
correction like e.g.  the gluon vacuum polarization (cf. $h_0^{QCD}$ above )
must be accounted for as a separate perturbation and must under no
circumstances be mixed with higher order leading logs in perturbation theory,
as summarized in a running $\a_s$. Our result provides a theoretical basis for
including only time-dilatation effects in $\G$ in explicit computations of the
Green function for toponium near threshold \pcite{kuehn}.

\vspace{.5cm}
One of the authors (W.M) thanks the organizers for financial support.  This
work is supported in part by the Austrian Science Foundation (FWF) in project
P10063-PHY within the framework of the EEC- Program "Human Capital and
Mobility", Network "Physics at High Energy Colliders", contract CHRX-CT93-0537
(DG 12 COMA).

\end{document}